\documentstyle[preprint,aps]{revtex}

\begin{document}
\title
{\large Persistent current for interacting
electrons: a simple Hartree-Fock picture}
\author{G. Montambaux}
\address{Laboratoire de Physique des
Solides,  associ\'e au CNRS \\
Universit\'{e} Paris--Sud \\ 91405 Orsay, France}
\maketitle
\centerline{June 13 1995}
\vspace*{1.4cm}
\begin{abstract}
The  average persistent
 current $\langle I \rangle$ of diffusive electrons in the Hartree-Fock
approximation is derived in a simple non-diagrammatic picture.
The Fourier expansion directly reflects the winding number
decomposition of the diffusive motion
around the ring. One recovers the results of Ambegaokar and
Eckern, and Schmid. Moreover one finds an expression for $\langle I \rangle$
which is valid beyond the diffusive regime.
\end{abstract}
\pacs{05.60+w, 72.10-d, 72.15Rn}
\vspace{1cm}
The physics of persistent currents in mesoscopic isolated rings pierced by
an Aharonov-Bohm flux $\phi$ , has attracted a lot of interest on the
description of the thermodynamic properties of mesoscopic metals, both in
the non-interacting picture or in the presence of electron-electron
interactions. The description of the interactions is a complicated task:
although several attempts have been made to describe the role of the
interactions in one-dimensional or few-channel rings, with the help of
analytical arguments or numerical calculations\cite{interactions}, the
diffusive nature of the
electronic motion, which is probably essential in the experiments on
metallic
rings with finite thickness\cite{Levy90,Chandrasekhar91,Mailly93}, has
been
treated only in two series of papers by Ambegaokar and Eckern and Schmid\cite
{Ambegaokar90,Schmid91}. They calculated the average persistent current in
the Hartree-Fock apprximation where the interaction is treated
perturbatively in a diagrammatic picture. Although this calculation provides
an average current smaller that experimentally observed\cite{Levy90}, one
can believe that it contains the essential physical ingredient namely weakly
interacting diffusive electrons. Thus it may be interesting to simplify as
much as possible the calculation in order to possibly generalize it, or to
compare it with numerical calculations. In the above papers, the average
persistent current is calculated with a diagrammatic technique where the
diffusive motion is described by a Cooperon pole with a wave vector
quantized
by periodic boundary conditions. Then, by Poisson summation, this sum over
diffusive modes is transformed into a sum over harmonics with periodicity $%
\phi _0/2$ where $\phi _0$ is the flux quantum.

In this short note, we propose a simple derivation where the average current
is directly related to the return probability for a diffusive particle.\
Then this return probability is expanded according to the winding number of
the diffusive motion around the ring, which gives directly access to the
harmonics expansion of the current.  Moreover, we obtain a general
expression which can be used beyond the diffusive regime.

Consider a quasi-one dimensional ring of perimeter $L$ and of transverse
section $S$, in which the motion is supposed to be diffusive along the
perimeter and uniform along the transverse section.

The first step is to write the total energy $E_T$ in the Hartree-Fock
approximation:
\begin{eqnarray}\label{hf3}
 E_T &=& E_T^0 +
 \sum_{i,j} \int U({\bf r}-{\bf r}') |\psi_j({\bf r}')|^2 |\psi_i({\bf r})|^2
d{\bf r}d{\bf r}'    \\
&-&\sum_{i,j}
\delta_{\sigma_i \sigma_j} \int U({\bf r}-{\bf r}')  \psi_j^{\ast}({\bf r}')
\psi_j({\bf r})
\psi_i^{\ast}({\bf r}) \psi_i({\bf r}') d{\bf r}d{\bf r}'      \nonumber
\end{eqnarray}
where $E_T^0$ is the total energy in the absence of interaction. In the
lowest order in the interaction parameter, the states
$\psi_i$ are the states of the {\it %
non-interacting} system. The summation $\sum_{i,j}$ is over filled
energy levels. $\sigma_i$ is the spin of a state $\psi_i$.

Considering that the Coulomb interaction $U({\bf r}-{\bf r}^{\prime })$ is
screened in the metallic regime, it is replaced by $U({\bf r}-{\bf r}%
^{\prime })=U\delta ({\bf r}-{\bf r}^{\prime })$ where $U=4\pi e^2/q_{TF}^2$
and $q_{TF}$ is the Thomas-Fermi wave vector. Replacing the interaction in
eq.\ref{hf3} by a $\delta $ function is certainly correct as long as the
Thomas-Fermi wave length is smaller than the mean free path $l_e$: $%
q_{TF}l_e\gg 1$. The interaction being now considered as $\delta $-like, it
is quite easy to see that the Fock term has the same structure as the
Hartree term. Introducing the local density $n({\bf r})=\sum_i|\psi _i({\bf r%
})|^2$, the total energy can be rewritten:
\begin{equation}
\label{hfscr}E_T=E_T^0+U\int n^2({\bf r})d{\bf r}-{\frac U2}\int n^2({\bf r}%
)d{\bf r}
\end{equation}
The Fock term is half the Hartree contribution because of the constraint on
the spin and its sign is opposite because of exchange of particles. The
local density $n({\bf r})$ can be expressed in terms of the Green function: $%
n({\bf r})=(-1/\pi )\int_0^{\epsilon _F}ImG^R({\bf r},{\bf r},\omega
)d\omega $ so that the average current is given by\cite{forget}
\begin{eqnarray}
\label{Iee}&&\langle I_{e-e}(\phi)\rangle =\langle -{\frac{\partial E_T}{
\partial \phi }}\rangle   \\
&=&-{\frac U{4\pi ^2}}{\frac \partial
{\partial \phi
}} \int_0^{\epsilon _F}\int_0^{\epsilon _F}\int \langle G^R({\bf r},{\bf r}
,\omega )G^A({\bf r},{\bf r},\omega ^{\prime })\rangle d\omega d\omega
^{\prime }d{\bf r}       \nonumber
\end{eqnarray}
where $G^R$ ($G^A$) is the retarded (advanced) Green function. The product $%
\langle G^R({\bf r},{\bf r})G^A({\bf r},{\bf r})\rangle $ simply expresses
the probability to go from some point ${\bf r}$ to itself \cite{Feynman}.
More precisely, for a particle at energy
$E$\cite{Chakravarty86,Prigodin94}:
 \begin{equation}\label{PEw}
P(E,\omega )={\frac 1{2\pi \rho _0}}\langle G^R({\bf r},{\bf r},E+\omega
/2)G^A({\bf r},{\bf r},E-\omega /2)\rangle
\end{equation}
is the Fourier transform of the return probability $P(E,t)$ after a time $t$%
:
\begin{equation} \label{PEt}
P(E,t)={\frac 1{2\pi }}\int P(E,\omega )e^{-i\omega t}d\omega  \nonumber
\end{equation}
Because of disorder average, this return probability is independent of the
position ${\bf r}$. The current can now be expressed directly in terms of
$P(E,t)$. Neglecting the energy dependence ($P(E,t)=P(t)$), one gets:
\begin{equation}
\label{IPt}\langle I_{e-e}(\phi)\rangle =-\Omega U\rho _0{\frac 2\pi }{\frac
\partial {\partial \phi }}\int_0^\infty {\frac{P(t,\phi )}{t^2}}dt
\end{equation}
$\Omega =LS$ is the volume. Since the relation (\ref{PEw}) is exact, the
expression (\ref{IPt}) is quite general and
is valid beyond the diffusive regime. In the classical approximation for the
diffusive regime ($l_e \ll L$), the diffusion
probability is the solution of a classical diffusion equation $D\Delta
P=\partial P/\partial t$ where $D$ is the diffusion coefficient taken here
at the Fermi level, $D=v_Fl_e/3$. It is given by
\begin{equation}
P_{cl}({\bf r},{\bf r}^{\prime },t)={\frac 1{(4\pi Dt)^{d/2}}}e^{-|{\bf r}-%
{\bf r}^{\prime }|^2/4Dt}
\end{equation}
In the geometry of a quasi-one-dimensional ring, the return probability can
thus be expanded according to the winding number of the diffusive motion:
\begin{equation}
P(t,\phi)={\frac 1{S\sqrt{4\pi Dt}}}\sum_me^{-{\frac{m^2L^2}{4Dt}}}[1+\cos
(4\pi m\varphi )]
\end{equation}
where $\varphi =\phi /\phi _0$. The second term, of importance here, results
from the phase interference between time-reversed paths in the
semi-classical approximation, each path
accumulating a phase $\pm 2 \pi m \varphi$\cite{Bergmann84,Chakravarty86}.
In zero flux,
the return probability is twice the classical one. This expansion according
to the winding number gives directly the Fourier decomposition of the
current:
\begin{equation}
\langle I_{e-e}(\phi)\rangle =\sum_mI_m\sin (4\pi m\varphi )  \nonumber
\end{equation}
with
\begin{equation}
I_m={\frac{8mU\rho _0}{\phi _0\sqrt{4\pi E_c}}}\int_0^\infty e^{-{\frac{m^2}{%
4E_ct}}}{\frac{dt}{t^{5/2}}}
\end{equation}
We have introduced the Thouless energy $E_c=\hbar D/L^2$. Defining a
dimensionless winding number $w$ by $w^2=m^2/4E_ct$, $I_m$ can be simply
rewritten as
\begin{eqnarray}
I_m &=& {64 \over \sqrt{\pi}}  {U \rho_0 \over \phi_0}{E_c \over
m^2}\int_0^\infty w^2 e^{-w^2} d w    \nonumber   \\
&=& 16  { U \rho_0 \over \phi_0} {E_c \over m^2}
\end{eqnarray}
which is the result of Ambegaokar and Eckern\cite{Ambegaokar90} and
Schmid\cite{Schmid91}.

We finish with the calculation of the current at finite temperature $T$
where two Fermi factors $f(\epsilon )$ have to be introduced in
eq.\ref{Iee}.
 Doing the standard substitution $\int f(\epsilon )g(\epsilon )d\epsilon
=2i\pi
\sum_{\omega _n}g(i\omega _n)$ where $\omega _n=(2n+1)\pi T$,
the average current at finite $T$ is
straightforwardly given by:
\begin{eqnarray}\label{IPtT}
\langle I_{e-e}\rangle &=&   - \Omega U \rho_0
{2 \over  \pi} {\partial \over \partial \phi} 4 \pi^2 T^2
\sum_{\omega_n,\omega_n'}
 P(i \omega_n- i\omega_n')  \nonumber \\
&=&   - \Omega U \rho_0
{2 \over  \pi} {\partial \over \partial \phi}
 \int  {\pi^2 T^2 \over ( \sinh \pi T t )^2} P(t,\phi) dt
\end{eqnarray}

By introducing the same dimensionless winding number $w$ as above, the
harmonics $I_m$ are given by
\begin{equation}
I_m(T)=I_m(0){\frac 4{\sqrt{\pi }}}\int_0^\infty dww^2e^{-w^2}({\frac{\pi
\theta _m}{4w^2}})^2{\frac 1{\sinh {}^2({\frac{\pi \theta _m}{4w^2}})}}
\end{equation}
where $T_m=E_c/m^2$ is the effective temperature associated with the winding
number $m$ and $\theta _m=T/T_m$. Although the integrand is quite
 different,
this temperature dependence is identical to the one found by Ambegaokar and
Eckern\cite{Ambegaokar90}. It directly  expresses the current in terms of a
temperature square average of a winding number.

In conclusion, we have calculated the average persistent current in the first
order of the Hartree-Fock approximation. By writing the current directly
in terms of the winding number decomposition of the return
probability, we have avoided the use of the diagrammatic
calculation and directly found the harmonic expansion of the
current. Although this calculation is reminiscent of the
semiclassical description of the spectral correlations and of
the average current of non-interacting particles in the
canonical ensemble\cite{Argaman93}, it does not
 use any semiclassical sum rule,
 since here the correlation function of interest can be exactly
written in terms of the return probability without any
approximation. Moreover, we have written an expression for the
average current (eqs.\ref{IPt},\ref{IPtT}) which is valid beyond the
diffusive regime and may be used for example in the ballistic regime
where interesting magnetic response have also been
observed\cite{Levy93}.


\newpage


\begin{thebibliography}{99}
\bibitem{interactions}A.
M\"uller-Groeling, H.A.
Weidenmuller and C.H. Lewenkopf,
Europhys. Lett. {\bf 22},193
(1993); M. Abraham and Berkovits,
Phys. Rev. Lett. {\bf 70}, 1509
(1993); G. Bouzerar, D. Poilblanc
and G. Montambaux, Phys. Rev. B
{\bf 49}, 8258 (1994); H. Kato
and D. Yoshioka, Phys. Rev. B
{\bf 50}, 4943 (1994); T.
Giamarchi and B.S. Shastry,
unpublished; M. Ramin, B. Reulet and H. Bouchiat, unpublished; M. Kamal,
Z.H. Musslimani and A. Auerbach, unpublished.

\bibitem{Levy90}L.P. L\'evy, G. Dolan, J. Dunsmuir and H. Bouchiat,
Phys.Rev. Lett.{\bf 64}, 2074 (1990).

\bibitem{Chandrasekhar91}V. Chandrasekhar, R.A. Webb, M.J. Brady,
M.B. Ketchen, W.J. Gallaghern and A. Kleinsasser, Phys.Rev.Lett.{\bf 67},
3578 (1991).

\bibitem{Mailly93}D. Mailly, C. Chapelier and A. Benoit, Phys.Rev.Lett.
{\bf  70}, 2020 (1993).


\bibitem{Ambegaokar90}V. Ambegaokar and U. Eckern, Phys. Rev. Lett. {\bf 65
}, 381 (1990);{\it ibid} {\bf 67}, 3192 (1991).

\bibitem{Schmid91}A. Schmid, Phys.Rev.Lett.{\bf 66}, 80 (1991).

\bibitem{forget}Other non flux dependent products of Green functions have
been omitted in this expression.



\bibitem{Feynman}R.P. Feynman and A.R. Hibbs: {\em Quantum--Mechanics and
Path--Integrals} (McGraw Hill, 1965);
 R.P. Feynman, Rev.Mod.Phys. {\bf 20}, 367 (1984).


\bibitem{Chakravarty86}S. Chakravarty and A. Schmid, Phys.Rep.{\bf 140}, 193
(1986).

\bibitem{Prigodin94}V.N. Prigodin, B.L. Altshuler, K.B. Efetov and S.
Iida, Phys. Rev. Lett. {\bf 72}, 546 (1994).

\bibitem{Bergmann84}G. Bergmann, Phys.Rep. {\bf 107}, 1 (1984).



\bibitem{Argaman93}N. Argaman, Y. Imry and U. Smilansky, Phys. Rev. B {\bf
47} 4440 (1993).

\bibitem{Levy93}L.P. L\'evy, D.H. Reich, L. Pfeiffer and K. West,
Physica (Amsterdam){\bf 189 B},204  (1994).

\end{thebibliography}
\end{document}